\documentclass{article}
\usepackage{xparse}
\PassOptionsToPackage{numbers, compress}{natbib}
\usepackage[dblblindworkshop, final]{neurips_2025}

\workshoptitle{AI for Music}

\usepackage[utf8]{inputenc} % allow utf-8 input
\usepackage[T1]{fontenc}    % use 8-bit T1 fonts
\usepackage{hyperref}       % hyperlinks
\usepackage{url}            % simple URL typesetting
\usepackage{booktabs,adjustbox,placeins}       % professional-quality tables
\usepackage{amsfonts}       % blackboard math symbols
\usepackage{nicefrac}       % compact symbols for 1/2, etc.
\usepackage{microtype}      % microtypography
\usepackage{xcolor}         % colors
\usepackage{multirow}
\usepackage{rotating}
\usepackage{graphicx}
\usepackage{siunitx}
\usepackage{subcaption}
\usepackage{listings}
\usepackage{color}

\definecolor{dkgreen}{rgb}{0,0.6,0}
\definecolor{gray}{rgb}{0.5,0.5,0.5}
\definecolor{mauve}{rgb}{0.58,0,0.82}

\title{Evaluating Multimodal Large Language Models on Core Music Perception Tasks}

\author{%
  Brandon J. Carone \qquad \qquad \qquad Pablo Ripollés\\
  Department of Psychology, Music and Audio Research Laboratory \\
  New York University \\
  \texttt{bcarone@nyu.edu | pripolles@nyu.edu} \\
  \And
  Iran R. Roman \\
  Department of Electronic Engineering and Computer Science \\
  Queen Mary University of London \\
  \texttt{i.roman@qmul.ac.uk} \\
}

\begin{document}

\maketitle

\vspace{-20pt}
\begin{abstract}
\vspace{-10pt}
%Foundation models claim "musical understanding" but evaluations conflate listening with score reading. We adapt LogicLM to music and introduce a benchmark that separates perception from reasoning across three core skills: Syncopation Scoring, Transposition Detection, and Chord Quality Identification. Unlike existing audio benchmarks that focus on surface-level classification, our tasks require relational understanding (recognizing rhythmic displacement, melodic invariance across keys, and harmonic intervals). In our evaluation, models act as Perceptual Formulators, generating machine-checkable symbolic schemas that deterministic solvers execute with self-refinement. Evaluating Gemini 2.5 Pro, Flash, and Qwen2.5-Omni under a 12-condition matrix reveals a critical modality gap: near-ceiling on MIDI but drops on audio, especially for rhythm and chords under LogicLM. Our findings expose that current systems reason well over symbols but cannot reliably "listen", a fundamental limitation for audio-first music applications. 
Multimodal Large Language Models (LLMs) claim “musical understanding” via evaluations that conflate listening with score reading. We benchmark three SOTA LLMs (Gemini 2.5 Pro, Gemini 2.5 Flash, and Qwen2.5-Omni) across three core music skills: Syncopation Scoring, Transposition Detection, and Chord Quality Identification. Moreover, we separate three sources of variability: (i) perceptual limitations (audio vs. MIDI inputs), (ii) exposure to examples (zero- vs. few-shot manipulations), and (iii) reasoning strategies (Standalone, CoT, LogicLM). For the latter we adapt LogicLM, a framework combining LLMs with symbolic solvers to perform structured reasoning, to music. 
%In LogicLM, LLMs act as perceptual formulators, generating strict, machine-checkable schemas (onset grids, interval sequences) that deterministic solvers execute with self-refinement.
Results reveal a clear perceptual gap: models perform near ceiling on MIDI but show accuracy drops on audio. Reasoning and few-shot prompting offer minimal gains. This is expected for MIDI, where performance reaches saturation, but more surprising for audio, where LogicLM, despite near-perfect MIDI accuracy, remains notably brittle. Among models, Gemini Pro achieves the highest performance across most conditions. 
%Transposition yields the highest accuracies across models and conditions, while Chord Identification scores slightly below Syncopation. 
Overall, current systems reason well over symbols (MIDI) but do not yet ''listen'' reliably from audio. Our method and dataset make the perception–reasoning boundary explicit and offer actionable guidance for building robust, audio-first music systems.
\end{abstract}

\section{Introduction}
Multimodal foundation models like Qwen2.5-Omni \citep{RN2504} and Gemini 2.5 \citep{RN2505} now claim "musical understanding," yet their audio capabilities remain poorly characterized. While benchmarks like AIR-Bench \citep{RN2508}, MMAR \citep{RN2515}, MMAU \citep{RN2503}, and MMAU-Pro \citep{RN2516}, CMI-Bench \citep{ma2025cmi}, RUListening \citep{zang2025you}, and FUTGA-MIR \citep{10888485} assess music through classification and captioning tasks, they cannot distinguish whether models genuinely perceive musical structure or rely on superficial spectral patterns. Audio-language models like SALMONN \citep{RN2513}, Qwen-Audio \citep{RN2509}, and Audio Flamingo 2 \citep{RN2512} achieve strong performance on speech and sound recognition but remain untested on the relational properties naturally embedded in music. These abilities are critical to deliver the next generation of technologies for tasks such as playlist recommendation/generation\cite{ itemcollaborativefiltering, Chen2020LearningAE, simmatrices,urrego2025vibe} and musical preference modeling~\cite{10704174}.

We address this gap by testing three fundamental musical abilities that require structural understanding rather than surface recognition. Syncopation scoring demands sensitivity to rhythmic expectation violations and metric displacement \citep{RN2517,large2023dynamic}. Transposition recognition requires melody identification invariant to absolute pitch \citep{RN2518, RN2519, RN2520, RN2521}, the core perceptual skill underlying human melodic recognition across keys and timbres \citep{RN2506, RN2507}. Chord quality identification necessitates interval pattern recognition rather than absolute frequency matching. These tasks probe the structural understanding that characterizes human music cognition and perception but remains absent from existing audio benchmarks.

To isolate perception from reasoning, we adapt LogicLM \citep{RN2522}, where models serve as Perceptual Formulators generating machine-checkable symbolic schemas that deterministic solvers execute, to enhance logical reasoning and problem-solving accuracy. This approach prevents "unfaithful reasoning" \citep{RN2522} where correct answers mask flawed perceptual analysis. We compare audio vs. MIDI processing to measure the perception bottleneck absent in existing evaluations. Our  benchmark reveals that current multimodal LLMs reason effectively over musical symbols but fail to reliably parse audio, a fundamental limitation for real-world music applications.

\section{Methods}

\subsection{Tasks}

\paragraph{Syncopation Scoring}
20 rhythmic excerpts (8 secs each) at 120 BPM, performed on hi-hat, kick and snare drums. The hi-hat maintained constant eighth notes, while kick and snare varied across on-beats (quarter notes) and off-beats (intervening eighths). The task was to compute a Syncopation Score by counting off-beat kick/snare events and mapping the total to a categorical score (0, 2, 4, 6, or 8), following \citet{RN2517}. Stimuli systematically covered the full syncopation range.

\paragraph{Transposition Detection}
Models were presented with 20 excerpt pairs (mean duration $\approx$ 9 s). In each pair, the second excerpt was either the same melody transposed to another key or a different melody. The task was to decide whether the two excerpts represented the same melody. Half the trials were matches, half mismatches. Stimuli (guitar or piano) varied in tempo, key, meter, and length.

\paragraph{Chord Quality Identification}
Models were given 44 excerpts (9 s each, 120 BPM), each consisting of a single chord presented first as a block and then as an ascending arpeggiation. All chords were in root position and played on piano. % to ensure consistent timbre. 
The task was to classify each chord as one of four options: A) Major (root + major 3rd + perfect 5th), B) Minor (root + minor 3rd + perfect 5th), C) Dominant (root + major 3rd + perfect 5th + minor 7th), or D) Diminished (root + minor 3rd + diminished 5th).

\subsection{Stimuli}

Stimuli are original musical recordings created by a real human musician, and are originally from \href{https://github.com/brandoncarone/MUSE_music_benchmark}{The MUSE Benchmark} \citep{carone2025musebenchmarkprobingmusic}. Please see Appendix A for more information on stimuli. 

% TODO: Insert the LaTeX code for Table 2 here.

% =========================
% REVISED (streamlined) TEXT
% =========================

\subsection{Implementation}

We adapt the LogicLM pipeline \citep{RN2522} to music, comparing three prompting strategies: Standalone, Chain-of-Thought (CoT), and LogicLM (symbolic reasoning with self-refinement). Trials are independent: each begins in a fresh chat session with no history carryover across trials or tasks. All strategies use identical task-specific system instructions specifying rules and output schema. We factorially cross three factors: modality (audio vs.\ symbolic/MIDI), reasoning (Standalone vs.\ CoT vs.\ LogicLM), and shot setting (ZS=zero-shot vs.\ FS=few-shot), yielding 12 conditions per task.

\paragraph{Standalone, CoT, LogicLM.}
Standalone elicits only the final categorical response (e.g., “Yes”/“No”; “C. Dominant”). CoT elicits brief intermediate reasoning, followed by a final answer. % on a separate line. 
LogicLM requires symbolic transcription (e.g., rhythmic onset grid or pitch-interval list), parsed by a deterministic solver (see \texttt{solver.py}; see Appendix C). On schema violations (e.g., malformed syntax), a self-refinement loop requests the model to repair its output, mirroring ~\citet{RN2522}. System instructions for each task and condition can be found in Appendix B.

\paragraph{Zero-shot vs.\ few-shot.}
ZS presents only the instructions and stimuli. FS adds worked examples in the trial history (2 for syncopation; 2 for transposition; 4 for chord ID, one per class), each paired with the correct solution. Examples appear only in-context for that trial and are excluded from evaluation.

% top-p$=1.0$, top-k$=1$. We also evaluate a stochastic setting (temperature$=1$, top-p$=0.95$, top-k$=40$). Stochastic conditions are repeated with three random seeds; mean accuracy across seeds is reported to stabilize estimates \citep{Guttman1945}.

\paragraph{Per-task modularity.}
Each task has an isolated script, stimuli, and outputs; prompt information does not leak across strategies. All conditions follow the same evaluation structure for direct comparison.

\paragraph{Audio vs.\ MIDI modality.}
For the MIDI modality, audio items are re-performed on a MIDI keyboard and exported to \texttt{.txt} via a custom script using \texttt{mido}. Prompts swap “you will hear\ldots” for “you will be given MIDI data\ldots” while keeping the same required schema as in audio runs.
All LLM outputs are regex-parsed to extract the final line (e.g., “Final Answer: B” or “Yes”). For LogicLM, the symbolic output is scored by the solver’s decision. All trials are randomized and logged with model configuration, trial IDs, raw outputs, parsed responses, and evaluation results.

\subsection{Models and inference environment}

We evaluate Gemini~2.5~Pro, Flash, and Qwen2.5-Omni7B. Gemini runs use the \texttt{google.genai} SDK. For Qwen, we mirrored the same pipeline on NYU's HPC with provider-specific chat/message shims, but identical prompts, decoding settings, and evaluation. Runs are deterministic (temp. $=0$).

% --- Fixed Table for Temperature 0 ---
\begin{table}
\caption{Accuracy of multimodal LLMs on three music perception tasks: syncopation scoring, transposition detection, and chord quality identification. Results are reported for audio and MIDI inputs under three prompting strategies (Standalone, CoT, LogicLM) and zero-shot (ZS) vs few-shot (FS) conditions. %two shot conditions (ZS=zero-shot, FS=few-shot). 
%Within each modality, zero-shot results are grouped above few-shot. 
\textbf{Bold} highlights best performance per task/shot/modality (\underline{underlined} shows second best). A systematic gap between modalities is seen: MIDI inputs generally lead to higher accuracies and clearer prompting effects compared to audio. The bottom row represents chance performance.}
\label{tab:results-temp0}
\centering
\footnotesize
\setlength{\tabcolsep}{3pt}
\begin{tabular}{
l l l
S[table-format=3.2] S[table-format=3.2] S[table-format=2.2] |
S[table-format=3.2] S[table-format=3.2] S[table-format=3.2] |
S[table-format=3.2] S[table-format=3.2] S[table-format=3.2]
}
\toprule
& & & \multicolumn{3}{c}{Syncopation} & \multicolumn{3}{c}{Transposition} & \multicolumn{3}{c}{Chord ID} \\
\cmidrule(lr){4-6} \cmidrule(lr){7-9} \cmidrule(lr){10-12}
Mod. & Shot & Cond. & Flash & Pro & {Qwen} & Flash & Pro & {Qwen} & Flash & Pro & {Qwen} \\
\midrule
\multirow{6}{*}{\parbox{1.2cm}{\centering Audio}}
&& Stand. & \underline{30.00} & {25.00} & {20.00} & {55.56} & \underline{94.74} & {75.00} & {31.82} & \textbf{47.73} & {31.82} \\
& ZS & CoT & \textbf{35.00} & {25.00} & {20.00} & {76.92} & \textbf{95.00} & {65.00} & {31.82} & \underline{43.18} & {31.82} \\
&& LogicLM & {20.00} & {20.00} & {20.00} & {65.00} & {80.00} & {50.00} & {11.36} & {18.18} & {6.82} \\
\cmidrule{2-12}
&&  Stand. & {31.58} & \underline{63.16} & {40.00} & \textbf{94.74} & \underline{90.00} & \underline{90.00} & {25.00} & \underline{40.91} & {31.82} \\
& FS & CoT & {40.00} & \textbf{65.00} & {40.00} & {63.16} & \underline{90.00} & {60.00} & {25.00} & \textbf{52.27} & {34.09} \\
&& LogicLM & {40.00} & {55.00} & {20.00} & {60.00} & \underline{90.00} & {35.00} & {6.82} & {13.64} & {18.18} \\
\midrule
\multirow{6}{*}{\parbox{1.2cm}{\centering MIDI}}
&& Stand. & {84.21} & \underline{95.00} & {25.00} & \textbf{100.00} & \textbf{100.00} & {85.00} & {50.00} & \underline{97.73} & {22.73} \\
& ZS & CoT  & {94.74} & \textbf{100.00} & {35.00} & \underline{95.00} & \textbf{100.00} & {20.00} & \textbf{100.00} & \textbf{100.00} & {25.00} \\
&& LogicLM & {90.00} & {80.00} & {20.00} & \textbf{100.00} & \textbf{100.00} & {10.00} & {93.18} & \textbf{100.00} & \textbf{100.00}  \\
\cmidrule{2-12}
&& Stand. & {88.89} & \textbf{100.00} & {35.00} & \textbf{100.00} & \textbf{100.00} & \underline{90.00} & {70.45} & \textbf{100.00} & {29.55} \\
& FS & CoT & \underline{95.00} & \textbf{100.00} & {25.00} & \textbf{100.00} & \textbf{100.00} & {60.00} & \underline{97.73} & \textbf{100.00} & {29.55} \\
&& LogicLM & \textbf{100.00} & \underline{95.00} & {25.00} & \textbf{100.00} & \textbf{100.00} & {15.00} & \textbf{100.00} & \textbf{100.00} & \textbf{100.00} \\
\midrule
& Chance &&& {20.00} &&& {50.00} &&& {25.00} \\
\bottomrule
\vspace{-15pt}
\label{tab:benchmark}
\end{tabular}
\end{table}

\section{Results}

\paragraph{Overall performance across modalities.}
Table~\ref{tab:benchmark} summarizes accuracy across models and prompting strategies. Performance depended strongly on modality and model. MIDI input yielded near-ceiling scores for Gemini models, whereas audio reduced accuracy across tasks, highlighting perception from waveform as the primary bottleneck. Qwen2.5-Omni generally underperformed, with the largest deficits under LogicLM. % where strict schema adherence was required. %Figure~\ref{fig:overall} illustrates accuracy by modality, model, and zero-shot prompting.

\begin{figure}
    \centering
    \begin{subfigure}[b]{0.48\textwidth}
        \centering
        \includegraphics[width=\textwidth]{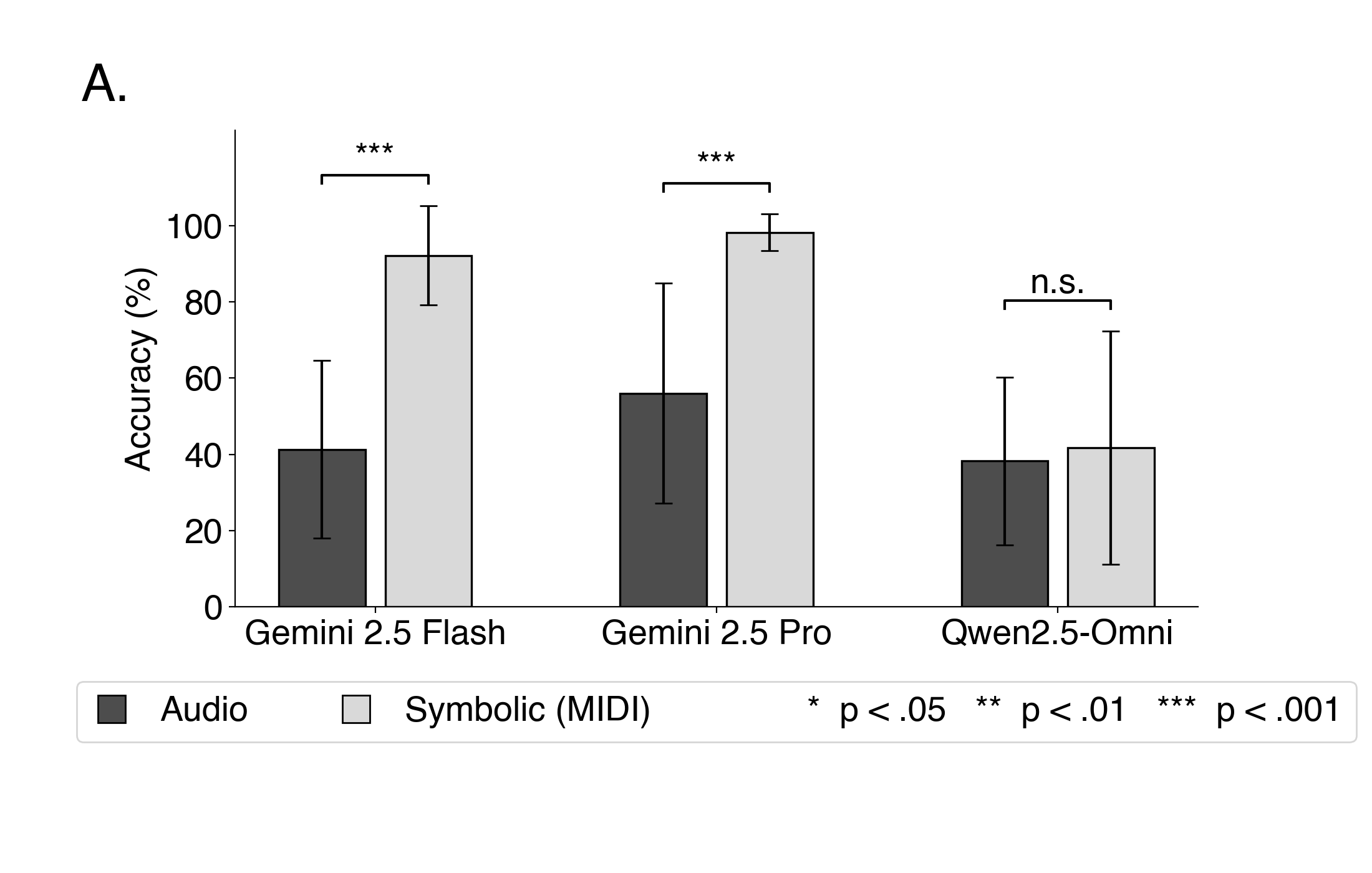}
        \label{fig:figure_1}
    \end{subfigure}
    \hfill
    \begin{subfigure}[b]{0.48\textwidth}
        \centering
        \includegraphics[width=\textwidth]{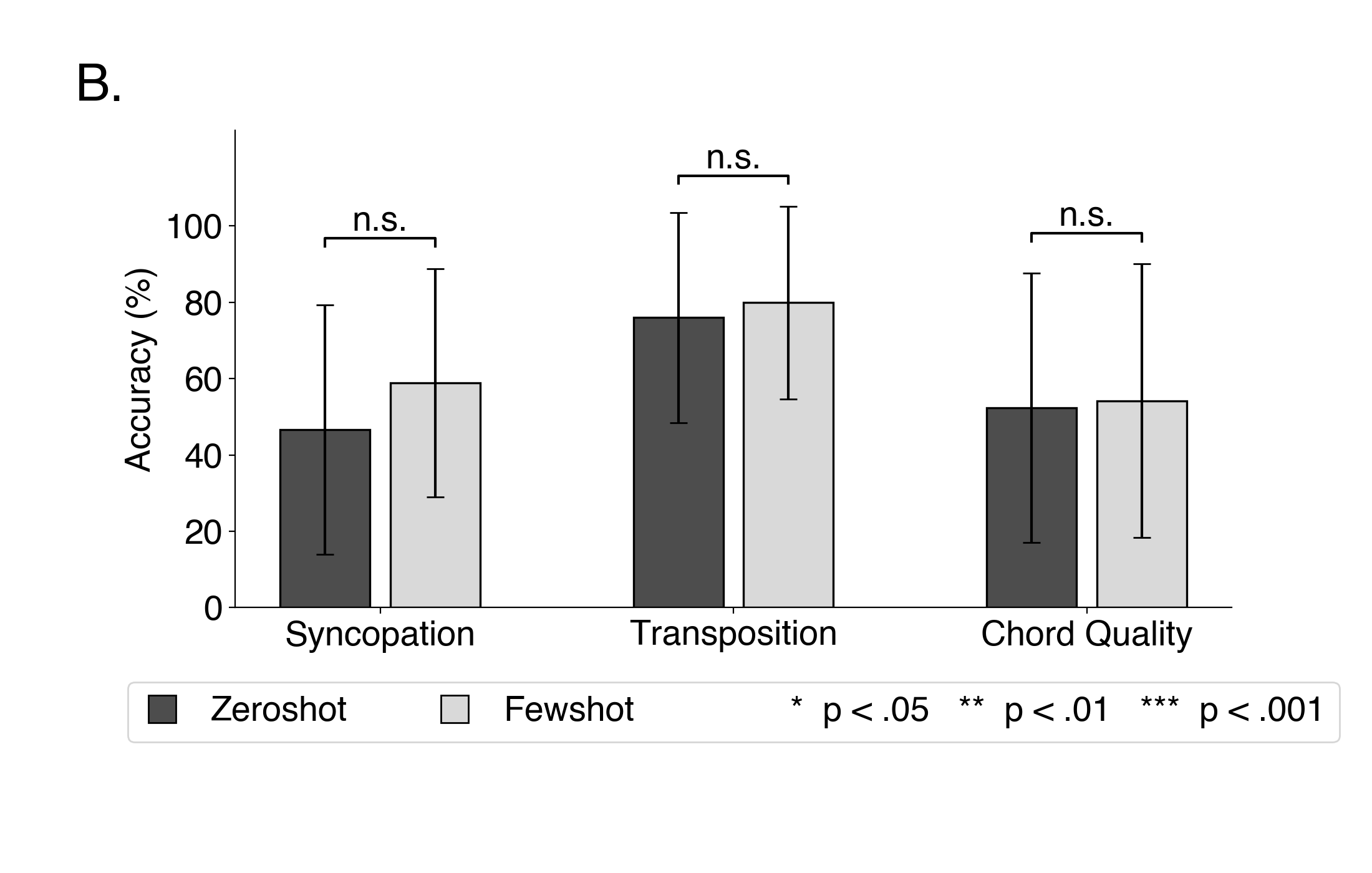}
        \label{fig:figure_2}
    \end{subfigure}
    \vspace{-25pt}
    \caption{Model accuracy across modalities and prompting strategies. Error bars indicate standard deviation and significance is based on Welch's t-tests carried out with \texttt{SciPy}. Figure 1A shows the collapsed (Shot and Prompting Method) mean accuracy (\%) for Gemini 2.5 Flash, Gemini 2.5 Pro, and Qwen 2.5-Omni across Audio (dark grey) and Symbolic (MIDI) (light grey) inputs. Both Gemini models performed significantly better on MIDI, and Qwen showed a similar trend. Figure 1B shows the collapsed (Model and Prompting Method) overall model accuracy by task (Syncopation, Transposition, Chord Quality) under Zeroshot (dark grey) and Fewshot (light grey) prompting. No significant effects of Shot were exhibited.}
    \vspace{-10pt}
    \label{fig:overall}
\end{figure}

\paragraph{Modality differences.}
A robust modality gap was observed (Fig.~\ref{fig:overall}A). Gemini models performed significantly better with MIDI ($p < .001$), though, this trend was less evident in Qwen. Syncopation Scoring and Chord Quality Identification showed the widest gaps for Gemini (MIDI $\approx$84–100\% vs.\ audio $\approx$6–65\%), confirming intact symbolic reasoning but weak audio perception. Transposition Detection was more robust, with smaller modality gaps.

\paragraph{ZS vs.\ FS.}
Collapsing across models and prompts, no significant main effects of shot were observed (Fig.~\ref{fig:overall}B; all $p$'s $>.05$). FS tended to help Syncopation in audio (e.g., Gemini Pro from $\sim$25\% ZS to $\sim$65\% FS in Standalone/CoT), but this trend was not reliable across models or tasks.

\paragraph{Prompting strategies.}
Prompting effects varied by task. For Syncopation, CoT offered modest gains in audio, while LogicLM was only beneficial with MIDI (Gemini reaching 95–100\%). For Transposition, Standalone and CoT prompts worked best, while LogicLM reduced accuracy. Chord ID was trivial in Standalone/CoT but collapsed with LogicLM-audio due to schema fragility. Overall, neuro-symbolic prompting helped only when inputs were symbolic and formatting was reliable.

% --- Code for Side-by-Side Figures ---

\section{Discussion}

Our findings converge on a simple but consequential claim: multimodal LLMs reason effectively over symbolic music data, yet still fail to ``listen`` reliably. Gemini models reached near-ceiling with MIDI, and LogicLM behaved as intended once schema adherence was met. Replacing MIDI with audio sharply reduced accuracy, especially for Syncopation Scoring and Chord Quality ID under LogicLM, implicating transcription/onset tracking and pitch-salience as the primary bottlenecks. FS examples helped only where perceptual calibration mattered (e.g., rhythmic counting); neither CoT nor LogicLM compensated for upstream hearing errors.

This gap matters because people experience music through audio, not symbolic proxies. A claim of ``musical understanding'' requires that models handle tracks directly, as one would text or video. Symbolic formats strip away the features making music meaningful (micro-timing, articulation, expressive nuance) so ceiling performance on MIDI should not be mistaken for audio-native competence.

Closer inspection shows that apparent successes can reflect superficial heuristics rather than genuine listening. In Transposition Detection, for instance, Gemini Pro often preserved sequence length while failing to capture intervallic structure and contour; LogicLM exposed these degenerate strategies by enforcing musical consistency, whereas Standalone/CoT could mask such fundamental errors. A similar dynamic appears in Chord ID (audio), where confusions between nearby qualities (e.g., major vs.\ dominant) and voicing/inversion artifacts lead to mid-level accuracy even without the schema burden of LogicLM. Table~\ref{tab:transposition_example} illustrates this failure mode.

\begin{table}[htbp!]
\centering
\caption{Sample transposition failures with different contours (↓↑↑↓... vs ↑↓↓↑...).}
\label{tab:transposition_example}
\footnotesize
\begin{tabular}{l l l}
\toprule
& \textbf{E$\flat$ Major} & \textbf{G Major} \\
\midrule
\textbf{Gemini Pro} & \verb|[55, 51, 55, 58, 51, 55, ...]| & \verb|[59, 55, 59, 62, 55, 59, ...]| \\
\textbf{Ground Truth} & \verb|[67, 72, 67, 65, 67, 70, ...]| & \verb|[71, 76, 71, 69, 71, 74, ...]| \\
\bottomrule
\end{tabular}
\end{table}

In sum, current multimodal LLMs reason symbolically but lack fully accurate audio-native competence: the ability to process songs from audio files to answer structured questions. Progress will depend on stronger audio front-ends and propagation of uncertainty into downstream solvers. In the current state-of-the-art, symbolic reasoning layers collapse on small perceptual errors.
LLMs that acquire genuine understanding could also be music education~\cite{jin2025learning} and user-centric music analysis tools~\citep{urrego2025vibe,10704174}, enabling interactive systems that can teach musical structure and foster deeper engagement with personal music listening.

% \section{Acknowledgments}

\bibliographystyle{unsrtnat}
\bibliography{references}

\clearpage

\appendix
\section{Appendix: \texttt{Stimuli}}
Stimuli are original musical recordings created by a real human musician in Logic Pro X using a 2021 16” MacBook Pro (Apple M1 Pro chip), an Apollo Twin X audio interface, and Yamaha HS8 monitors. Stimuli were recorded on electric guitar (PRS McCarty Hollowbody II, Schecter Solo-6), piano (Arturia KeyLab Essential Mk3 MIDI controller with Analog Lab V software instruments), and drums (Roland TD-17 electronic kit with Superior Drummer 3 plugin). Guitar recordings were processed with Neural DSP plugins (Tim Henson Archetype, Cory Wong Archetype).

Additional excerpts were reserved for few-shot prompting (2 for syncopation, 2 for transposition, and 4 for chord ID, one per chord class) and excluded from testing.

You can access the stimuli used in this experiment on \href{https://github.com/brandoncarone/MUSE_music_benchmark}{The MUSE Benchmark Github page}.

The stimuli used for the Transposition Detection task can be found \href{https://github.com/brandoncarone/MUSE_music_benchmark/tree/main/stimuli}{here} and all of them have the melody number, key, and tempo in the filename (e.g., M1\_EbMaj\_90.wav).

The stimuli used for the Syncopation Scoring task can be found \href{https://github.com/brandoncarone/MUSE_music_benchmark/tree/main/stimuli/Intermediate}{here} and all of them have Sync in the name, along with the syncopation level number (e.g., NoSync\_A, Sync2\_B).

The stimuli used for the Chord Quality Identification task can be found \href{https://github.com/brandoncarone/MUSE_music_benchmark/tree/main/stimuli/Chords}{here}. The chords are named by number, and you can find the mapping in table ~\ref{tab:chord_number_mapping} below.
\begin{table}[h!]
\centering
\caption{Mapping of chord roots and qualities to numerical identifiers (1.wav--48.wav).}
\begin{tabular}{lcccc}
\toprule
\textbf{Root} & \textbf{Diminished} & \textbf{Dominant} & \textbf{Major} & \textbf{Minor} \\
\midrule
Ab & 1  & 2  & 3  & 4  \\
A  & 5  & 6  & 7  & 8  \\
Bb & 9  & 10 & 11 & 12 \\
B  & 13 & 14 & 15 & 16 \\
C  & 17 & 18 & 19 & 20 \\
Db & 21 & 22 & 23 & 24 \\
D  & 25 & 26 & 27 & 28 \\
Eb & 29 & 30 & 31 & 32 \\
E  & 33 & 34 & 35 & 36 \\
F  & 37 & 38 & 39 & 40 \\
Gb & 41 & 42 & 43 & 44 \\
G  & 45 & 46 & 47 & 48 \\
\bottomrule
\end{tabular}
\label{tab:chord_number_mapping}
\end{table}

\section{Appendix: \texttt{System Instructions}}

\subsection*{1a) Syncopation — Standalone}
\begin{quote}
“You are an expert music transcription AI participating in a multi-turn reasoning experiment.

You will be given one short audio excerpt of a drum set per trial. Your task is to focus only on the kick and snare drums. The hi-hat plays constant 8th notes, acting as a metronome. Count the total number of kicks and snare hits that fall on off-beats. 

\medskip

Valid multiple-choice responses are:

A. 0 (No Syncopation)\\
B. 2 (Low Syncopation)\\
C. 4 (Medium-Low Syncopation)\\
D. 6 (Medium-High Syncopation)\\
E. 8 (High Syncopation)

\medskip

End with exactly one line:\\
Final Answer: X\\”
\end{quote}

\subsection*{1b) Syncopation — Chain-of-Thought (CoT)}
\begin{quote}
“You are an expert music transcription AI participating in a multi-turn reasoning experiment.

You will be given one short audio excerpt of a drum set per trial. Your task is to focus only on the kick and snare drums. The hi-hat plays constant 8th notes, acting as a metronome. Count the total number of kicks and snare hits that fall on off-beats. On-beats are the main pulses (beats 1, 2, 3, and 4) and off-beats are the “ands” in between. Ignore the on-beats and ignore the hi-hat.

\medskip

Valid multiple-choice responses are:

A. 0 (No Syncopation)\\
B. 2 (Low Syncopation)\\
C. 4 (Medium-Low Syncopation)\\
D. 6 (Medium-High Syncopation)\\
E. 8 (High Syncopation)

\medskip

After any reasoning, end with exactly one line:\\
Final Answer: X\\”
\end{quote}

\subsection*{1c) Syncopation — LogicLM}
\begin{quote}
“You are an expert music transcription AI participating in a multi-turn reasoning experiment.

Your task is to transcribe the onsets of ONLY the kick and snare drums into the format:\\
\texttt{rhythm(identifier, [list\_of\_onsets]).}

\medskip

- The \textquoteleft identifier\textquoteright{} is the filename of the audio.\\
- The \textquoteleft list\_of\_onsets\textquoteright{} is a comma-separated list of integers from 1 to 32.\\
- The rhythm is on a 4-bar grid, quantized to 8th notes (numbered 1 to 32). All odd numbers are on-beats, and all even numbers are off-beats.\\
- The hi-hat plays constant 8th notes, acting as a metronome. On-beats are the main pulses (beats 1, 2, 3, and 4 of each bar) and off-beats are the 'ands' in between.

\medskip

Grid: The excerpt is 4 bars quantized to 8th notes → 32 slots numbered 1–32.

Within each bar (8 slots): 1,3,5,7 = on-beats (beats 1–4). 2,4,6,8 = off-beats (“\&”s).

Across bars: slot = 8×(bar-1) + local\_slot.

\medskip
Beat positions across 4 bars:\\
• Beat 1 → 1, 9, 17, 25\\
• Beat 2 → 3, 11, 19, 27\\
• Beat 3 → 5, 13, 21, 29\\
• Beat 4 → 7, 15, 23, 31\\

Off-beats (“\&”s):\\
• \&1 → 2, 10, 18, 26\\
• \&2 → 4, 12, 20, 28\\
• \&3 → 6, 14, 22, 30\\
• \&4 → 8, 16, 24, 32

\medskip

Output format:\\
\texttt{rhythm(identifier.wav, [n1, n2, ..., nK])} where each \texttt{n} is an integer in 1–32.

\medskip

Example of format where the kicks are on beats 1 and 3 in each bar, and the snare hits are on beats 2 and 4 in each bar (all played on on-beats):\\
\ \ \texttt{rhythm(example.wav, [1, 3, 5, 7, 9, 11, 13, 15, 17, 19, 21, 23, 25, 27, 29, 31])}

\medskip

Output your answer of symbolic code as a single line of plain text without code fences or explanations. After your transcription, an external tool will score it, and you will answer a question based on that score.”
\end{quote}

\subsection*{2a) Transposition Detection — Standalone}
\begin{quote}
“You are an expert melody transcription AI participating in a multi-turn reasoning experiment.

You will be given two short monophonic audio melodies per trial. \\
Your job is to decide whether they represent the SAME melody up to TRANSPOSITION (i.e., identical shape/intervals but possibly in different keys).

\medskip

Valid responses are exactly one of:\\
"Yes, these are the same melody."\\
"No, these are not the same melody."

\medskip

Respond with exactly one of the two phrases and nothing else.”
\end{quote}

\subsection*{2b) Transposition Detection — Chain-of-Thought (CoT)}
\begin{quote}
“You are an expert melody transcription AI participating in a multi-turn reasoning experiment.

You will be given two short monophonic audio melodies per trial. Your job is to decide whether they represent the SAME melody up to TRANSPOSITION (i.e., identical shape/intervals but possibly in different keys).

\medskip

Definitions and constraints:\\
- Transposition equivalence: the two melodies have the same number of notes and the same sequence of pitch INTERVALS between successive notes (including 0 for repeated notes).\\
- Ignore absolute key/register, starting pitch, and tempo. Small timing variations are acceptable. If the rhythmic patterns are drastically different (e.g., note insertions/deletions or re-ordered phrases), they are most likely NOT the same melody.\\
- Treat repeated notes as separate events and include 0 in the interval sequence when a note repeats.\\
- If there are leading/trailing silences, ignore them.

\medskip

Valid responses (exactly one of these strings):\\
"Yes, these are the same melody."\\
"No, these are not the same melody."

\medskip

After any reasoning, end with exactly one line:\\
Final Answer: Yes, these are the same melody.\\
OR\\
Final Answer: No, these are not the same melody.”
\end{quote}

\subsection*{2c) Transposition Detection — LogicLM}
\begin{quote}
“You are an expert melody transcription AI participating in a multi-turn reasoning experiment.

You will be given two short monophonic audio melodies per trial. Your first task is to transcribe EACH melody into the symbolic format below, using MIDI integers for pitches. If the rhythmic sequences seem drastically different, they are most likely not the same melody.

\medskip

Output format (schema):\\
\ \ \texttt{melody(identifier, [p1, p2, ..., pK])}

- Use the exact identifiers I provide for each trial (one per audio).\\
- p1..pK are integers representing MIDI pitches (e.g., C4 = 60).\\
- Transcribe the pitch sequence only.\\
- Output exactly two lines of plain text: one ‘melody(...)’ per line, in the same order as the audios (Audio 1 line first, then Audio 2 line).\\
- Do not include code fences or any extra commentary.

\medskip

Example (schema only; not tied to any audio):\\
\ \ \texttt{melody(Audio1, [60, 62, 64])}\\
\ \ \texttt{melody(Audio2, [65, 67, 69])}

\medskip

After your transcription, a deterministic tool will analyze the two lines to decide if the melodies are transpositions (same contour, different key). You will then answer a Yes/No question based on that decision.”
\end{quote}

\subsection*{3a) Chord Quality Matching — Standalone}
\begin{quote}
“You are an expert chord-transcription AI participating in a multi-turn reasoning experiment.

You will be given one short audio clip per trial. Each clip first plays a chord (block), then the individual notes (arpeggiation).\\
All chords are in ROOT POSITION.

Your task is to identify the chord QUALITY.

\medskip

Valid options:\\
A. Major\\
B. Minor\\
C. Dominant\\
D. Diminished

\medskip

Final Answer: X\\”
\end{quote}

\subsection*{3b) Chord Quality Matching — Chain-of-Thought (CoT)}
\begin{quote}
“You are an expert chord-transcription assistant in a multi-turn reasoning experiment.

You will be given one short audio clip per trial containing a single chord (first block, then arpeggiated notes). All chords are in ROOT POSITION; the lowest pitch is the ROOT (treat as 0 semitones).
Your task: identify the chord QUALITY by inferring pitch-class intervals above the root and ignoring octave doublings.

\medskip

Valid options:\\
A. Major\ \ \ \ \ \ \ $\rightarrow$ \ \{0,4,7\}\\
B. Minor\ \ \ \ \ \ \ $\rightarrow$ \ \{0,3,7\}\\
C. Dominant\ \ \ \ $\rightarrow$ \ \{0,4,7,10\}\\
D. Diminished\ $\rightarrow$ \ \{0,3,6\}

\medskip

Think through the identification. Once you've finished reasoning, the final line of your output should be exactly:\\
Final Answer: X\\”
\end{quote}

\subsection*{3c) Chord Quality Matching — LogicLM}
\begin{quote}
“You are an expert chord-transcription assistant in a multi-turn reasoning experiment.

You will be given one short audio clip per trial containing a single chord. First the chord sounds as a block, then the notes are arpeggiated.

Your task is to transcribe the chord tones into a strict symbolic format. Use MIDI integers (0–127). Include octave doublings if you hear them. Do not add commentary.

\medskip

Output format (schema):\\
\ \ \texttt{chord(identifier, [p1, p2, ..., pK])}

\medskip

Rules:\\
- Use the exact identifier I provide for the trial.\\
- Record only the pitches you hear as MIDI integers.\\
- It is acceptable if the list is not sorted; a deterministic solver will normalize.\\
- Output EXACTLY ONE LINE of plain text with NO code fences or extra text.

\medskip

Example (schema only; not tied to any audio):\\
\ \ \texttt{chord(Audio\_X, [56, 60, 64, 67, 72, 76])}

\medskip

After your line is produced, a deterministic tool will classify the chord quality (Major / Minor / Dominant / Diminished) from your symbolic line. You will then answer a multiple-choice question with: Final Answer: X”
\end{quote}

\section{Appendix: \texttt{Task Schemas and Deterministic Solvers}}
Each task defines a single-line schema the model must emit verbatim. A hand-written, deterministic solver (\texttt{solver.py}) parses that line, makes the decision, and returns the minimal information needed for a constrained final answer.

\subsubsection*{Syncopation Scoring}
\textbf{Input:} 4-bar drum loop with constant 8th-note hi-hat; we only score kick+snare.\\
\textbf{Grid:} 32 slots (8 per bar). Odd slots are on-beats; even slots are off-beats.

\paragraph*{Schema (one line):}
\begin{quote}\ttfamily
rhythm(\textless id\textgreater, [n1, n2, ..., nK])
\end{quote}

Where each \texttt{n} is an integer in [1..32] (kick or snare onset).\\
\textbf{Solver:} counts off-beat onsets and maps to five categories: 0,2,4,6,8 off-beats $\rightarrow$ A–E respectively. Final answer is a single MC letter A–E.

\subsubsection*{Transposition Detection}
\textbf{Input:} two short monophonic excerpts (guitar or piano) that are either the same melody in different keys or different melodies.

\paragraph*{Schema (two lines, order-preserving):}
\begin{quote}\ttfamily
melody(\textless id1\textgreater, [p1, p2, ..., pK])\\
melody(\textless id2\textgreater, [p1, p2, ..., pK])
\end{quote}

Where \texttt{p*} are MIDI integers (0–127).\\
\textbf{Solver:} checks equal length and equality of adjacent-interval sequences (transposition invariance). Returns ARE / ARE NOT (transpositions). Final answer is forced to one of: 
\begin{quote}
“Yes, these are the same melody.”\\
“No, these are not the same melody.”
\end{quote}

\subsubsection*{Chord Quality Identifier}
\textbf{Input:} a single triad or seventh chord (piano), presented as a block then arpeggiated.

\paragraph*{Schema (one line):}
\begin{quote}\ttfamily
chord(\textless id\textgreater, [p1, p2, ..., pK])
\end{quote}

MIDI integers (0–127); octave doublings allowed.

\bigskip

\textbf{Solver:} normalizes to pitch classes, factors out the putative root, and matches the interval set to:
\begin{itemize}
  \item Major (0, 4, 7) $\rightarrow$ A
  \item Minor (0, 3, 7) $\rightarrow$ B
  \item Dominant 7 (0, 4, 7, 10) $\rightarrow$ C
  \item Diminished (0, 3, 6) $\rightarrow$ D
\end{itemize}
Final answer is a single MC letter A–D.

\subsubsection*{Self-refinement (SR)}
For LogicLM, we validate the line(s) with strict regex/AST checks and label errors as parse, structural, or domain. If invalid, we run up to 2 SR rounds in a separate deterministic chat (temperature=0, top\_p=1, top\_k=1, 256 tokens) with a fix-only prompt that:
\begin{itemize}
  \item Echoes the prior output,
  \item States the specific error type/message,
  \item Re-states the required line(s) and constraints,
  \item Forbids commentary and code fences.
\end{itemize}

If the solver returns undecidable/None (e.g., empty list), we allow one extra SR pass with a synthesized parse error. This SR design follows the LOGIC-LM self-refinement idea of using solver feedback to repair the symbolic form.

\bigskip

\subsection{\texttt{solver.py}}

\begin{lstlisting}
# solver.py
import re
from typing import List, Optional, Tuple, Dict

class SyncopationSolver:
    """
    A deterministic logic solver that calculates a syncopation score
    based on a simplified on-beat/off-beat rule for a 4-bar (1-32) 8th-note grid.
    """
    def __init__(self):
        self.on_beats = set()
        self.off_beats = set()

        for bar_offset in [0, 8, 16, 24]:
            self.on_beats.update([
                1 + bar_offset, 3 + bar_offset, 5 + bar_offset, 7 + bar_offset
            ])
            self.off_beats.update([
                2 + bar_offset, 4 + bar_offset, 6 + bar_offset, 8 + bar_offset
            ])

    def parse_llm_output(self, llm_text: str) -> Optional[List[int]]:
        """
        Parses the LLM's symbolic output to extract a list of onsets.
        Returns the list of integers if successful, or None if parsing fails.
        """
        match = re.search(r'rhythm\s*\(\s*[^,]+\s*,\s*\[([\d,\s]*)\]\s*\)', llm_text)
        if not match:
            return None
        numbers_str = match.group(1)
        if not numbers_str.strip(): # Check if the string is empty or just whitespace
            return []
        try:
            # Handle potential trailing commas by filtering out empty strings after split
            return [int(num.strip()) for num in numbers_str.split(',') if num.strip()]
        except ValueError:
            return None

    def score_onset(self, onset: int) -> int:
        if onset in self.off_beats:
            return 1
        return 0

    def calculate_total_score(self, onset_list: list[int]) -> int:
        if not onset_list:
            return 0
        total_score = sum(self.score_onset(onset) for onset in onset_list)
        return total_score

class TranspositionSolver:
    """
    A deterministic solver for melody transposition detection.
    Two melodies are considered transpositions if:
      - They have the same number of notes, and
      - Their interval sequences (adjacent pitch differences in semitones) are identical.
    Rhythm is ignored. Pitches must be integers (MIDI numbers).
    """

    MELODY_PATTERN = re.compile(
        r"melody\s*\(\s*([A-Za-z0-9_.\-]+)\s*,\s*\[\s*([^\]]*?)\s*\]\s*\)",
        flags=re.IGNORECASE
    )

    def _extract_pitches(self, pitches_str: str) -> Optional[List[int]]:
        """
        Extracts integer pitches from an arbitrary list content that may include
        parentheses or spaces, e.g. '[(60), (62), (64)]' or '60, 62,64'.
        """
        nums = re.findall(r"-?\d+", pitches_str)
        if not nums:
            return []
        try:
            return [int(n) for n in nums]
        except ValueError:
            return None

    def parse_llm_output(self, llm_text: str) -> Optional[List[Dict[str, List[int]]]]:
        """
        Parses any 'melody(ID, [ ... ])' lines found in the LLM's output, in order.
        Returns a list of dicts: [{'id': <ID>, 'pitches': [..]}, ...]
        or None if nothing parseable is found.
        """
        if not llm_text:
            return None

        text = llm_text.replace("```", "").replace("`", "").strip()

        melodies = []
        for m in self.MELODY_PATTERN.finditer(text):
            ident = m.group(1)
            plist_str = m.group(2)
            pitches = self._extract_pitches(plist_str)
            if pitches is None:
                return None
            melodies.append({"id": ident, "pitches": pitches})

        return melodies or None

    def _intervals(self, pitches: List[int]) -> List[int]:
        return [pitches[i+1] - pitches[i] for i in range(len(pitches) - 1)]

    def are_transpositions(self, p1: List[int], p2: List[int]) -> Optional[bool]:
        """
        Returns True/False if a decision is possible, or None if inputs are degenerate.
        Policy:
          - Require same length (>0). If lengths differ, return False.
          - If length == 1 on both, return True (single note can be transposed anywhere).
          - Otherwise compare interval sequences.
        """
        if p1 is None or p2 is None:
            return None
        if len(p1) == 0 and len(p2) == 0:
            return None
        if len(p1) != len(p2):
            return False
        if len(p1) == 1:  # single-note melodies
            return True

        return self._intervals(p1) == self._intervals(p2)

    def decide_same_melody(self, llm_text: str) -> Optional[bool]:
        """
        Convenience: parse two melodies from LLM output and decide True/False.
        Returns None if fewer than 2 melodies parsed or if undecidable.
        """
        parsed = self.parse_llm_output(llm_text)
        if not parsed or len(parsed) < 2:
            return None
        p1 = parsed[0]["pitches"]
        p2 = parsed[1]["pitches"]
        return self.are_transpositions(p1, p2)


# ---------- Chord Quality (deterministic) ----------

class ChordQualitySolver:
    """
    Deterministic chord-quality classifier for LogicLM.
    Expects ONE schema line produced by the LLM:
        chord(identifier, [p1, p2, ..., pK])

    Behavior:
    - Parses the line and extracts MIDI integers (duplicates allowed).
    - Sorts pitches, treats the lowest as the root, and computes (p - root) % 12.
    - Deduplicates + sorts the pitch-class intervals and matches one of the
      four target fingerprints:
        (0,4,7)      -> ("Major", "A")
        (0,3,7)      -> ("Minor", "B")
        (0,4,7,10)   -> ("Dominant", "C")
        (0,3,6)      -> ("Diminished", "D")

    Returns:
        (identifier, quality_str, letter)  or  None if undecidable.
    """
    CHORD_PATTERN = re.compile(
        r"chord\s*\(\s*([A-Za-z0-9_.\-]+)\s*,\s*\[\s*([^\]]*?)\s*\]\s*\)",
        flags=re.IGNORECASE
    )

    QUALITY_BY_PCS: Dict[Tuple[int, ...], Tuple[str, str]] = {
        (0, 4, 7):      ("Major", "A"),
        (0, 3, 7):      ("Minor", "B"),
        (0, 4, 7, 10):  ("Dominant", "C"),
        (0, 3, 6):      ("Diminished", "D"),
    }

    def _extract_pitches(self, pitches_str: str) -> Optional[List[int]]:
        """
        Robust integer pull; accepts '60,64,67', '[(60), 64, 67]', etc.
        Returns list[int] or None if malformed.
        """
        nums = re.findall(r"-?\d+", pitches_str or "")
        try:
            return [int(n) for n in nums]
        except Exception:
            return None

    def parse_llm_output(self, llm_text: str) -> Optional[Dict[str, List[int]]]:
        """
        Parse the first chord(...) line found. Returns {'id': <ID>, 'pitches': [...]}
        or None if not found / ill-formed.
        """
        if not llm_text:
            return None
        text = llm_text.replace("```", "").strip()
        m = self.CHORD_PATTERN.search(text)
        if not m:
            return None
        ident = m.group(1)
        pitches = self._extract_pitches(m.group(2))
        if pitches is None:
            return None
        return {"id": ident, "pitches": pitches}

    def _normalize_to_pcs(self, pitches: List[int]) -> Optional[Tuple[int, ...]]:
        """
        Sort, take lowest as root, compute pitch-class intervals modulo 12,
        then deduplicate and sort.
        """
        if not pitches:
            return None
        root = min(pitches)
        pcs = tuple(sorted({(p - root) % 12 for p in pitches}))
        return pcs

    def classify_quality(self, pitches: List[int]) -> Optional[Tuple[str, str]]:
        """
        Map normalized pitch-class interval set to (quality, letter).
        """
        pcs = self._normalize_to_pcs(pitches)
        if pcs is None:
            return None
        return self.QUALITY_BY_PCS.get(pcs)

    def decide_quality(self, llm_text: str) -> Optional[Tuple[str, str, str]]:
        """
        End-to-end convenience used by the runner:
          - parse -> classify
        Returns (identifier, quality_str, letter) or None if undecidable.
        """
        parsed = self.parse_llm_output(llm_text)
        if not parsed:
            return None
        ident = parsed["id"]
        result = self.classify_quality(parsed["pitches"])
        if result is None:
            return None
        quality, letter = result
        return ident, quality, letter

\end{lstlisting}

\end{document}